\documentstyle[11pt]{article}\textheight 230mm\textwidth 150mm
            \newcommand{\be}{\begin{eqnarray}}
            \newcommand{\ee}{\end{eqnarray}}
            \newcommand{\eel}[1]{\label{#1}\end{eqnarray}}
\newcommand{\e}[1]{\label{e:#1}\end{eqnarray}}
     \newcommand{\eg}{{\em e.g.\ }}
            \newcommand{\ie}{{\em i.e.\ }}

            \newcommand{\la}{{\lambda}}
            
            \newcommand{\del}{{\delta}}

           \newcommand{\ra}{{\rightarrow}}
 \newcommand{\lea}{{\leftarrow}}
            
            \newcommand{\Lra}{{\Leftrightarrow}}
            \newcommand{\pet}{{\cal P}}

\newcommand{\ca}{{\cal C}}

            \newcommand{\beq}{\begin{quote}}
            \newcommand{\eq}{\end{quote}}
            \newcommand{\Om}{\Omega}
   
            \newcommand{\al}{\alpha}
            \newcommand{\ben}{\begin{enumerate}}
            \newcommand{\een}{\end{enumerate}}
            \newcommand{\bit}{\begin{itemize}}
            \newcommand{\ei}{\end{itemize}}
    	\newcommand{\nn}{\nonumber}
            \newcommand{\r}[1]{(\ref{e:#1})}
            \newcommand{\edfl}[1]{\label{#1}\end{df}}
\def\theequation{\thesection.\arabic{equation}}
\newcommand{\vb}{{\cal h}}
\newcommand{\hb}{{\cal i}}

\newcommand{\ve}{{\varepsilon}}

\newcommand{\dagg}{^{\dag}}

\newcommand{\bett}{{\bf 1}}
	
\def\d{\partial}
\def\cC{{\cal C}}

  \def\half{{1 \over 2}}

\begin{document}
\begin{titlepage}
\noindent
G\"{o}teborg ITP 99-06\\

\vspace*{5 mm}
\vspace*{35mm}
\begin{center}{\LARGE\bf Open group  transformations}
\end{center} \vspace*{3 mm} \begin{center} \vspace*{3 mm}

\begin{center}Igor Batalin\footnote{On leave of absence from
P.N.Lebedev Physical Institute, 117924  Moscow, Russia\\E-mail:
batalin@td.lpi.ac.ru.} and Robert
Marnelius\footnote{E-mail: tferm@fy.chalmers.se.}\\
\vspace*{7 mm} {\sl Institute
of Theoretical Physics\\ Chalmers University of
Technology\\ G\"{o}teborg
University\\ S-412 96  G\"{o}teborg, Sweden}\end{center}
\vspace*{25 mm}
\begin{abstract}
Open groups whose generators are in arbitrary involutions may be quantized
within a
ghost extended framework in terms of a nilpotent BFV-BRST charge operator.
Previously
we have shown that generalized quantum Maurer-Cartan equations for
arbitrary open
groups may be extracted from the quantum connection operators and that they also
follow from a simple quantum master equation involving an extended
nilpotent BFV-BRST charge and a master charge. Here we give further
details of these results. In addition we establish the general structure of the
solutions of the quantum  master equation. We  also construct an
extended formulation whose properties are determined by the extended BRST
charge in
the master equation.
\end{abstract}\end{center}\end{titlepage}

\setcounter{page}{1}
\section{Introduction}
Lie groups are  dominant continuous transformations
in theoretical  physics. However,
there are important models in which more general kinds of   continuous
transformations enter. For instance, the gauge
transformations in field theories like
gravity, supergravity and p-branes are not
Lie groups. Open groups whose generators
are in arbitrary involutions are very
difficult to handle. The only framework in which
they can be treated systematically  in quantum
theory is within a ghost extended BRST
frame. This
 is consistent when the group algebra may be
imbedded in a nilpotent BRST charge
following the BFV-prescription \cite{BFV}.
Representations of open groups are
then naturally classified in terms of ranks
\cite{BFV}. Ordinary nonabelian Lie groups are of rank 1. The gauge group of
supergravity has rank 2
\cite{FV}, and that of p-branes rank p \cite{MH}.

In \cite{OG} we investigated the properties of group transformed operators for
arbitrary open groups within the BFV-BRST-scheme
\cite{BFV}. By means of a particular
representation of the operator connections in
the Lie equations for the transformed
operators, we arrived at generalized
Maurer-Cartan equations expressed in terms
of the quantum antibracket introduced in
\cite{Quanti}. Furthermore, we presented
a new type of quantum master equation, also
expressed in terms of the quantum antibracket, which we
proposed to encode these generalized
Maurer-Cartan equations. Here we give further
details of these results. In addition we
establish the general structure of the
solutions of the quantum  master equation.
We  also construct an
extended formulation whose properties are
determined by the extended BRST charge in
the master equation.

\section{Quantizing arbitrary involutions}
Consider a dynamical system with a
finite number of degrees of freedom. Let the phase
space (or the symplectic manifold), $\Gamma$,
for this system be spanned by the
coordinates $z^A$. Thus, $\{z^A, z^B\}$
is an invertable matrix in terms of the
Poisson bracket. (Locally $z^A$ consists of
canonical conjugate pairs: $q^i, p_i$.)
The generators of the open group under
consideration are functions on $\Gamma$
denoted by $\theta_a(z)$ with  Grassmann
parities
$\ve(\theta_a)\equiv\ve_a (=0, 1)$  satisfying the Lie algebra
 \be
 &&\{\theta_a(z), \theta_b(z)\}=U_{ab}^{\;\;\;c}(z)\theta_c(z).
\e{1}
As is well-known it is extremely
difficult to construct a quantum theory in which the
corresponding commutator algebra to \r{1} is integrable when the structure
coefficients $U_{ab}^{\;\;\;c}$ are
functions on $\Gamma$. The canonical procedure
which makes this possible is to
first extend the phase space $\Gamma$ by ghost
variables and then imbed the algebra
\r{1} in one odd real function
$\Om$ satisfying  $\{\Om, \Om\}=0$ in terms of the
extended Poisson bracket.
 $\Om$
is the BFV-BRST charge \cite{BFV}.
The corresponding quantum theory
is now consistent if the corresponding odd, hermitian operator
$\Om$ is nilpotent, \ie $\Om^2=0$.
For a finite number of degrees of freedom such a
solution always exists and is of
the form \cite{BF} ($N$ is the rank of the theory.
$\theta_a$ are assumed to be hermitian  operators.)
\be
 &&\Om=\sum_{i=0}^N\Om_i,
 \e{2}
\be
&&\Om_0\equiv\ca^a\theta_a(z),\;\;\;
\Om_i\equiv \Om_{a_1\cdots a_{i+1}}^{b_i\cdots
b_1}(z)(\pet_{b_1}\cdots\pet_{b_i}
\ca^{a_{i+1}}\cdots\ca^{a_1})_{Weyl},
\;\;\;i=1,\ldots,N,\nn\\
\e{3}
where $\cC^a$, $\pet_a$ are the ghost
operators with Grassmann parities
$\ve(\cC^a)=\ve(\pet_a)=\ve_a+1$, satisfying the properties
 \be
&&[\ca^a,
\pet_b]=i\hbar\del^a_b,\quad(\cC^a)\dagg=\cC^a,
\quad\pet_a\dagg=-(-1)^{\varepsilon_a}\pet_a.
\e{4}
All commutators are from now on  graded commutators defined by
\be
&&[A, B]\equiv AB-BA(-1)^{\ve_A\ve_B},
\e{41}
where $\ve_A$ and $\ve_B$ are the
Grassmann parities of the operators $A$ and $B$
respectively. In \r{3} the ghost
operators are Weyl ordered which means that $\Om_i$
are all hermitian.
 $\Om$ determines  the
precise form of the quantum
counterpart of the algebra \r{1}. A convenient form of
this algebra is obtained if we rewrite $\Om$ in the
 following $\cC\pet$-ordered form
\cite{BF}
 \be
 &&\Om=\sum_{i=0}^N\Om'_i,\;\;\;
\Om'_0\equiv\ca^a\theta'_a(z),\nn\\
&&\Om'_i\equiv \ca^{a_{i+1}}\cdots
\ca^{a_1}{\Om'}_{a_1\cdots a_{i+1}}^{b_i\cdots
b_1}(z)\pet_{b_1}\cdots\pet_{b_i},\;\;\;i=1,\ldots,N.
 \e{51}
The  nilpotency of $\Om$ requires then the commutator algebra
\be
&&[\theta'_a(z), \theta'_b(z)]=
i\hbar{U'}_{ab}^{\;\;\;c}(z)\theta'_c(z),
\e{6}
where the structure operators
${U'}_{ab}^{\;\;\;c}(z)$ are given by
\be
&&{U'}_{ab}^{\;\;\;c}(z)=2(-1)^{\ve_b+\ve_c}{\Om'}_{ab}^{\;\;\;c}(z).
\e{7}
In terms of the coefficient operators in \r{3},
${\Om'}_{ab}^{\;\;\;c}(z)$ and
$\theta'_a(z)$ are given by
\be
&&{\Om'}_{ab}^c(z)={\Om}_{ab}^c(z)
+\half\sum_{n=1}^\infty\left({i\hbar\over 2}
\right)^n(n+1)(n+2)!
\,\Om_{aba_1\cdots
a_n}^{a_n\cdots a_1 c}(z)(-1)^{\sum_{k=1}^n\ve_{a_k}},\nn\\
&&\theta_a'(z)=\theta_a(z)+\sum_{n=1}^\infty\left({i\hbar\over
2}\right)^n(n+1)!\,\Om^{a_n\cdots a_1}_{aa_1\cdots a_n}(z)
(-1)^{\sum_{k=1}^n\ve_{a_k}},
\e{8}
which shows that  $\theta'_a(z)$
in general are different from $\theta_a(z)$ and  in
general not even hermitian. However, the main point here is that
$\Om$  through \eg
\r{6} represents the quantum counterpart of
\r{1}.

All operators and states may be decomposed
into operators and states with definite
ghost numbers. The ghost number $g$ is defined by
\be
&&G|A\hb_g=i\hbar g|A\hb_g, \quad [G, A_g]=i\hbar g A_g,
\e{9}
where $G$ is the hermitian ghost charge operator defined by
\be
&&
G\equiv-\half\left(\pet_a\cC^a-\cC^a\pet_a(-1)^{\ve_a}\right).
\e{10}
One may notice that $\Om$ in \r{2} has ghost number one, \ie
$[G, \Om]=i\hbar\Om$.

In a BRST-quantization which requires
us to solve the BRST cohomology resulting from
the BRST condition
\be
&&\Om|phys\hb=0,
\e{11}
the original generators $\theta_a$ in \r{1}
are constraint variables which generate
gauge transformations. Within the BRST
quantization the gauge generators have a
BRST exact form, \ie they are of the
form $[\Om, \rho]$. The natural gauge generators
are
$[\Om,
\pet_a]$. However, $[\Om, \pet_a]$ do
not satisfy a closed algebra for higher rank
theories ($N=2$ and higher). On the other hand,
$[\Om, \pet_a]$ and $\pet_a$ which
constitute BRST-doublets do always satisfy a closed algebra.

\setcounter{equation}{0}
\section{Finite group transformations for open groups}

We want now to integrate the quantum
involution \r{1} encoded in $\Om$ as
represented by \eg \r{6}.  We consider
therefore the  Lie equations for the
group transformed states and operators given by
(cf \cite{OG})
\be
&&\vb
A(\phi)|\stackrel{\lea}{D}_a\equiv\vb
A(\phi)|\left(\stackrel{\lea}{\d_a}-
(i\hbar)^{-1} Y_a(\phi)\right)=0,
\e{301}
\be
&&A(\phi)\stackrel{\lea}{\nabla}_a\equiv
A(\phi)\stackrel{\lea}{\d_a}-(i\hbar)^{-1}
[A(\phi), Y_a(\phi)]=0,
\e{302}
where $\d_a$ is a derivative with
respect to the group parameter $\phi^a$,
$\ve(\phi^a)=\ve_a$. The connection operator
$Y_a$, which depends on $\phi^a$, must
satisfy integrability conditions
\be
Y_a\stackrel{\lea}{\d_b}-Y_b
\stackrel{\lea}{\d_a}(-1)^{\ve_a\ve_b}=
(i\hbar)^{-1}[Y_a,
Y_b].
\e{303}
A formal solution is
\be
&&Y_a(\phi)\equiv i\hbar U(\phi)
\left(U^{-1}(\phi)\stackrel{\lea}{\d_a}\right),
\e{304}
which means that the group transformed states and operators are of the form
\be
&&\vb A(\phi)|=\vb A|U^{-1}(\phi), \quad A(\phi)=U(\phi)AU^{-1}(\phi),
\e{305}
where $U(\phi)$ is a finite group
element. $U(\phi)$ must be an even operator
and it is natural to require it to satisfy the conditions
\be
&&U(0)=\bett, \quad [\Om, U(\phi)]=0, \quad [G, U(\phi)]=0.
\e{306}
For an exponential representation this implies
\be
&&U(\phi)=\exp{\{{i\over\hbar}F(\phi)\}},
\quad F(0)=0, \quad [\Om, F(\phi)]=0, \quad
[G, F(\phi)]=0.
\e{307}
The first condition is just a
choice of parametrization. The second condition is
necessary in order to be consistent
with a BRST quantization defined by \r{11}. The
last condition  makes the group
transformed state to have the same ghost
number as the original one. Since in a
BRST quantization the gauge generators are
represented by BRST exact operators it is natural to expect
$U(\phi)$ to have the form
\be
&&U(\phi)=\exp{\{-(i\hbar)^{-2}[\Om, \rho(\phi)]\}},
\e{308}
where $\rho(\phi)$ has ghost number minus one. It could \eg be
$\rho(\phi)\propto\pet_a\phi^a$.
(\r{307} requires $[\Om, \rho(0)]=0$.) It is of
course natural for
$U(\phi)$ to be a unitary operator.

Let us now go back to the Lie
equations \r{301} and \r{302}. The conditions
\r{306} applied to the formal expression
\r{304} for the connection operator $Y_a$ implies that
\be
&&[\Om, Y_a]=0, \quad [G, Y_a]=0.
\e{309}
In terms of the exponential
representation \r{307} we have also
\be
&&Y_a(\phi)=\int_0^1 d\al
\exp{\{{i\over\hbar}\al F(\phi)\}}
\left(F(\phi)\stackrel{\lea}{\d_a}\right)
\exp{\{-{i\over\hbar}\al F(\phi)\}}.
\e{310}
In particular we have
$\left.Y_a(0)=F(\phi)\stackrel{\lea}
{\d_a}\right|_{\phi=0}$. From the natural
representation \r{308} with
$\rho(\phi)=\pet_a\phi^a$ we have then
$Y_a(0)=[\Om, \pet_a]$ which are the
natural gauge generators.

The  representation \r{308} yields the general property
\be
&&Y_a(\phi)=(i\hbar)^{-1}[\Om, \Om_a(\phi)],
\quad\ve(\Om_a)=\ve_a+1,
\e{311}
where  $\Om_a$ has ghost number minus one.
In fact,  \r{308}  suggests that
\be
&&\Om_a(\phi)=\int_0^1 d\al
\exp{\{-(i\hbar)^{-2}\al[\Om,
\rho(\phi)]\}}\left(\rho(\phi)\stackrel{\lea}{\d_a}\right)
\exp{\{(i\hbar)^{-2}\al[\Om,
\rho(\phi)]\}}.\nn\\
\e{312}
The simplest ansatz $\rho(\phi)=\pet_a\phi^a$
implies that $\Om_a(\phi)$ has
the form
\be
&&\Om_a(\phi)=\la^b_a(\phi)\pet_b+\{\mbox{\small possible ghost
dependent terms}\},\quad
\la^b_a(0)=\del^b_a,
\e{313}
where $\la^b_a(\phi)$ are operators in general.
Notice that $\Om_a(\phi)$  is only
defined up to BRST invariant operators from a
given $Y_a(\phi)$ exactly like
$\rho(\phi)$ from a given $F(\phi)$. However,
since the only BRST invariant
operators with negative ghost numbers are BRST
exact ones the arbitrariness is
\be
&&\Om_a(\phi)\,\ra\,\Om_a(\phi)+
(i\hbar)^{-1}[\Om, K_a(\phi)], \quad
\rho(\phi)\,\ra\,\rho(\phi)+
(i\hbar)^{-1}[\Om, \kappa(\phi)],
\e{314}
where $K_a(\phi)$ and $\kappa(\phi)$
have ghost number minus two.
If the relation \r{311} is applied
to the exponential representation
\r{307} then we get from \r{310}
\be
&&F(\phi)\stackrel{\lea}{\d_a}=(i\hbar)^{-1}
[\Om, G_a(\phi)]\;\Rightarrow\;F(\phi)=
F(0)+(i\hbar)^{-1}[\Om, \rho(\phi)],
\e{3141}
where
\be
&&\rho(\phi)=\int_0^1 d\al G_a(\al\phi)\phi^a
+(i\hbar)^{-1}[\Om, \kappa(\phi)].
\e{3142}
(Contract the first relation in \r{3141}
 with $\phi^a$ and use the argument in
appendix B.) The operators $G_a(\phi)$ must
satisfy the integrability conditions
\be
&&G_a(\phi)\stackrel{\lea}{\d_b}-G_b(\phi)
\stackrel{\lea}{\d_a}(-1)^{\ve_a\ve_b}=(i\hbar)^{-1}[\Om,
G_{ab}(\phi)],
\e{3143}
where $G_{ab}(\phi)$ in turn must satisfy the
integrability conditions following from
\r{3143} and so on.
 Thus, if $F(0)=0$ as required in \r{307} then
the representation \r{308} follows from \r{311}.
Since the original algebra
\r{1} is the algebra of the  gauge group
in a BRST quantization and this is the
algebra we want to integrate  the representation \r{311} of the
connection operator
$Y_a(\phi)$ is the natural one.  The relation
\r{311} was also the starting
point in
\cite{OG}. There we noticed that the
integrability conditions \r{303} lead to a whole
set of integrability conditions for
$\Om_a(\phi)$. The representation \r{311}
inserted into
\r{303} implies
\be
&&[\Om, \Om_a\stackrel{\lea}{\d_b}-\Om_b
\stackrel{\lea}{\d_a}(-1)^{\ve_a\ve_b}-
(i\hbar)^{-2}(\Om_a, \Om_b)_{\Om}]=0,
\e{315}
where we have introduced the quantum
antibracket defined in
\r{a1} in appendix A. (They were
introduced in \cite{Quanti,GenQuanti} and the
relevant formulas are given in appendix A.)
Since the right entry has ghost number
minus one, it is zero up to a BRST exact
operator. We have therefore
\cite{OG}
\be
&&\Om_a\stackrel{\lea}{\d_b}-\Om_b
\stackrel{\lea}{\d_a}(-1)^{\ve_a\ve_b}-
(i\hbar)^{-2}(\Om_a, \Om_b)_{\Om}+\half
(i\hbar)^{-1}[\Om_{ab}, \Om]=0,
\e{316}
where $\Om_{ab}$ in general is a $\phi^a$-dependent
 operator with ghost number minus
two. From \r{316} one may then derive
integrability conditions for $\Om_{ab}$ which
in turn introduces an operator $\Om_{abc}$
with ghost number minus three, when the
$\Om$-commutator is divided out. Thus,  $Y_a$ is
replaced by a whole set of operators, and  the
integrability condition \r{303} for
$Y_a$ is replaced by a whole set of
integrability conditions for these operators.
 These integrability conditions may be
viewed as generalized Maurer-Cartan
equations. In fact, for Lie groups we may choose
\be
&&\Om_a=\la^b_a(\phi)\pet_b,\quad
[\Om_a, \Om_b]=0, \quad \Om_{ab}=0,
\e{317}
where $\la^b_a(\phi)$ only depends on $\phi^a$.
In this case \r{316} reduces to
\be
&&\d_a\la_b^c-\d_b\la_a^c(-1)^{\ve_a\ve_b}
=\la^e_a\la^d_b
U^c_{de}(-1)^{\ve_b\ve_e+\ve_c+\ve_d+\ve_e},
\quad \la^b_a(0)=\del^b_a,
\e{318}
which are the classical Maurer-Cartan equations.
In fact, \r{317} is also valid for
quasigroup first rank theories \cite{OG}.
We expect a nonzero $\Om_{ab}$ to be
necessary only for theories of rank
two and higher. That a weaker form of
Maurer-Cartan equations is necessary for
rank two and higher should be connected to
the fact that $[\Om, \pet_a]$ then no longer
satisfy a closed algebra.
However, notice that a nonzero $\Om_{ab}$
is possible even in Lie group theories due
to the ambiguity \r{314}. The Maurer-Cartan
equations \r{316} only retain their form
under the replacement \r{314} if we at
the same time make the replacement
\be
&\Om_{ab}\;\ra&\Om_{ab}+\biggl(2(K_a\stackrel{\lea}{\d_b}-K_b
\stackrel{\lea}{\d_a}(-1)^{\ve_a\ve_b})-
\nn\\&&-(i\hbar)^{-2}((K_a, \Om_b)_{\Om}-(K_b,
\Om_a)_{\Om}(-1)^{\ve_a\ve_b})\biggr)(-1)^{\ve_a+\ve_b}.
\e{319}

\setcounter{equation}{0}
\section{The quantum master equation}
In \cite{OG} we proposed a quantum
master equation for all the integrability
conditions for the $\Om$-operators that
follow from \r{303} and \r{311}. All the
$\Om$-operators were there imbedded in
one master charge $S$. It is given by
\be
&&S(\phi,
\eta)\equiv G+\eta^a\Om_a(\phi)+
\half\eta^b\eta^a\Om_{ab}(\phi)(-1)^{\ve_b}+
\nn\\&&+{1\over6}\eta^c\eta^b\eta^a\Om_{abc}
(\phi)(-1)^{\ve_b+\ve_a\ve_c}+
\ldots\nn\\&&\ldots+
{1\over n!}\eta^{a_n}\cdots\eta^{a_1}\Om_{a_1\cdots
a_n}(\phi)(-1)^{\ve_n}+\ldots,\nn\\&&
\ve_n\equiv \sum_{k=1}^{[{n\over
2}]}\ve_{a_{2k}}+\sum_{k=1}^{[{n-1\over
2}]}\ve_{a_{2k-1}}\ve_{a_{2k+1}},
\e{401}
where $G$ is the ghost charge operator defined
in \r{10}. The variables $\eta^a$,
$\ve(\eta^a)=\ve_a+1$, are new parameters
which may be viewed as superpartners to
$\phi^a$. The general sign factor $(-1)^{\ve_n}$,
 which is different from the one in
\cite{OG}, makes the $\Om$-operators
have the symmetry properties
\be
&&n \mbox{\ even\ or\ odd}:\quad
\Om_{\cdots a_{2k}a_{2k+1}\cdots}=-\Om_{\cdots
a_{2k+1}a_{2k}\cdots}(-1)^{\ve_{a_{2k-1}}
\ve_{a_{2k}}+\ve_{a_{2k}}\ve_{a_{2k+1}}
+\ve_{a_{2k+1}}\ve_{a_{2k-1}}},\nn\\
&&n \mbox{\ even}:\quad \Om_{\cdots a_{2k+1}
a_{2k+2}\cdots}=-\Om_{\cdots
a_{2k+2}a_{2k+1}\cdots}(-1)^{\ve_{a_{2k-1}}
\ve_{a_{2k+1}}+\ve_{a_{2k+1}}\ve_{a_{2k+2}}
+\ve_{a_{2k+2}}\ve_{a_{2k-1}}},\nn\\
&&n \mbox{\ odd}:\quad \Om_{\cdots a_{2k-1}
a_{2k}\cdots}=-\Om_{\cdots
a_{2k}a_{2k-1}\cdots}(-1)^{\ve_{a_{2k-1}}
\ve_{a_{2k}}+\ve_{a_{2k}}\ve_{a_{2k+1}}
+\ve_{a_{2k+1}}\ve_{a_{2k-1}}}.
\e{402}
In addition the $\Om$-operators satisfy
 the properties
\be
&&\ve(\Om_{a_1\cdots a_n})=\ve_{a_1}+
\ldots+\ve_{a_n}+n,\quad[G, \Om_{a_1\cdots
a_n}]=-n i\hbar\Om_{a_1\cdots a_n}.
\e{403}
The last relation implies that
$\Om_{a_1\cdots a_n}$ has ghost number minus $n$.
If we assign ghost
number one to $\eta^a$,  generalizing
the original ghost number, then
$S$ has ghost number zero.
The main conjecture in \cite{OG} was that
the integrability conditions \r{316}
together with those that follow from \r{316}
are contained in the quantum master
equation
\be
&&(S, S)_{\Delta}=i\hbar[\Delta, S],
\e{404}
where $\Delta$ is  the extended nilpotent
BFV-BRST charge operator given by
\be
&&\Delta\equiv\Om-i\hbar\eta^a \d_a,\quad \Delta^2=0,
\e{405}
$\Delta$ is odd and has extended ghost number one.
The quantum antibracket $(S, S)_{\Delta}$
is defined by \r{a1} in appendix A with $Q$
replaced by
$\Delta$. In terms of commutators it is
\be
&&\quad (S, S)_{\Delta}\equiv[[ S, \Delta], S].
\e{406}
The master equation \r{404} may therefore be written as
\be
&&[S, [ S, \Delta]]=i\hbar[S, \Delta].
\e{407}
The explicit form of $[S, \Delta]$ is
to the lowest orders in $\eta^a$
\be
&&[S, \Delta]=i\hbar\Om+\eta^a[\Om_a,
\Om]+\eta^b\eta^a\Om_a\stackrel{\lea}{\d_b}i\hbar(-1)^{\ve_b}+
\half\eta^b\eta^a[\Om_{ab},
\Om](-1)^{\ve_b}+\nn\\&&+\half\eta^c\eta^b\eta^a\Om_{ab}
\stackrel{\lea}{\d_c}i\hbar(-1)^{\ve_b+\ve_c}+{1\over
6}\eta^c\eta^b\eta^a[\Om_{abc},
\Om](-1)^{\ve_b+\ve_a\ve_c}+O(\eta^4).
\e{408}
When this is inserted into \r{407} we
find that the quantum master equation
is satisfied identically
to zeroth and first order in $\eta^a$. To
second order in $\eta^a$ it yields exactly \r{316}, and
 to the  third order in $\eta^a$ it yields
\be
&&\d_a\Om_{bc}(-1)^{\ve_a\ve_c}+\half(i\hbar)^{-2}(\Om_a,
\Om_{bc})_{\Om}(-1)^{\ve_a\ve_c}+
cycle(a,b,c)=\nn\\&&=-(i\hbar)^{-3}(\Om_a,
\Om_b,
\Om_c)_{\Om}(-1)^{\ve_a\ve_c}-{2\over3}
(i\hbar)^{-1}[\Om'_{abc},\Om],\nn\\
&&
\Om'_{abc}\equiv\Om_{abc}-{1\over8}
\left\{(i\hbar)^{-1}[\Om_{ab},
\Om_c](-1)^{\ve_a\ve_c}+cycle(a,b,c)\right\},
\e{409}
where
we have introduced the higher quantum
antibracket of order 3 defined by  \r{a9} in
appendix A with $Q$ replaced by $\Om$,
or equivalently by \r{a12} in terms of
the 2-antibracket. In fact, \r{409} coincides
with the integrability conditions of
\r{316}. (Notice that if $\Om_{ab}=\Om_{abc}=0$,
which we may have for rank one
theories, $\Om_a$ must satisfy the
condition $(\Om_a,
\Om_b,\Om_c)_{\Om}=0$. That this is
satisfied we have checked up to quasigroups
\cite{OG}.)

The master equation \r{407} requires for consistency
\be
&&[\Delta, (S, S)_{\Delta}]=0\quad
\Lra\quad [\Delta, S]^2=0.
\e{410}
This property was also checked in
\cite{OG} up to the third order in $\eta^a$. To the
second order it yields exactly
\r{303}.

Formally we may introduce conjugate momenta,
$\pi_a$ and $\xi_a$, to the variables
$\phi^a$ and $\eta^a$. We have then
\be
&&[\phi^a, \pi_b]=i\hbar\del^a_b,
\quad [\eta^a, \xi_b]=i\hbar\del^a_b.
\e{411}
All derivatives with respect to $\phi^a$
may then be replaced by $\pi_a$. The Lie
equations \r{301}-\r{302} may \eg be written as
\be
&&\vb A(\phi)|(\pi_a-Y_a(\phi))=0,\quad
[A(\phi), \pi_a-Y_a(\phi)]=0,
\e{412}
and the $\Delta$-operator \r{405} may be written as
\be
&&\Delta\equiv\Om+\eta^a \pi_a(-1)^{\ve_a}.
\e{413}
(This form was used in \cite{OG}.) The use of
$\xi_a$ is so far unclear since we have
not used any derivatives with respect to $\eta^a$.
However, it allows us to define a
total ghost charge by
\be
&&\tilde{G}\equiv G-\half\left(\xi_a\eta^a-
\eta^a\xi_a(-1)^{\ve_a}\right),
\e{414}
where $G$ is the ghost charge \r{10}.
In terms of $\tilde{G}$ we have
\be
&&[\tilde{G}, S]=0, \quad [\tilde{G},
\Delta]=i\hbar\Delta.
\e{415}

\setcounter{equation}{0}
\section{Formal properties of the quantum master equation}
Let us define the transformed operators
$S(\al)$ and $\Delta(\al)$ by
\be
&&S(\al)\equiv e^{{i\over\hbar}\al F}
Se^{-{i\over\hbar}\al F}, \quad
\Delta(\al)\equiv e^{{i\over\hbar}\al F}
{\Delta} e^{-{i\over\hbar}\al F},
\e{501}
where $\al$ is a real parameter and
$F$  an arbitrary even operator with total ghost
number zero since we require $S(\al)$ and
$\Delta(\al)$ to have total ghost number zero and one. If
$S$ and
$\Delta$ satisfy the master equation
\r{404}, then  it is easily seen that $S(\al)$
and
$\Delta(\al)$ satisfy the transformed master equation
\be
&&(S(\al), S(\al))_{\Delta(\al)}=
i\hbar[\Delta(\al), S(\al)].
\e{502}
If we restrict $F$ in \r{501}  to satisfy
the master equation \r{404}, \ie
\be
&&(F, F)_{\Delta}=i\hbar[\Delta, F],
\e{503}
then $\Delta(\al)$ in \r{501} reduces to
\be
&&\Delta(\al)=\Delta+(i\hbar)^{-1}
[\Delta, F](1-e^{-\al}).
\e{504}
(Notice that \r{503} implies $\Delta''(\al)+\Delta'(\al)=0$.)
 For $F=S$ we have then
that $S$ satisfies the master equation
\r{404} with $\Delta$ replaced by
$\Delta(\al)$ in \r{504} where $F=S$.

There are also transformations on $S$
leaving $\Delta$ unaffected for which the
master equation \r{404} is invariant.
From \r{501} this is the case if
$\Delta(\al)=\Delta$ which requires
\be
&&[\Delta, F]=0.
\e{505}
In order for the transformed $S$ to be in the form \r{401}, $F$ should
not depend on $\pi_a$ and $\xi_a$ in \r{411}. If we assume that
$F(\phi, \eta)$ may be given by a power series in $\phi^a$ and
$\eta^a$ we find that \r{505} implies (the proof is given in appendix B)
\be
&&F(\phi, \eta)=F_0+(i\hbar)^{-1}[\Delta, \Psi(\phi, \eta)],
\e{506}
where $F_0=F(0,0)$ and $\left.[\Delta,
\Psi(\phi, \eta)]\right|_{\phi=\eta=0}=0$.
$\Psi$ is an odd operator with total
ghost number minus one which does not depend on
$\pi_a$ and $\xi_a$. (It has the form
\r{c10} in appendix B.) An interesting
consequence of the result \r{506} is
that the extended BRST singlets are identical to
the original BRST singlets under $\Om$ since
$[\Delta, F_0]=[\Om, F_0]=0$. Now although
the general invariance transformation on
$S$ is given by \r{501} with the
$F$-operator \r{506}, it is natural and
consistent to set $F_0=0$ in which case
we find the following class of invariance transformations \cite{Sp2QA}
\be
&&S\;\ra\;S'\equiv \exp{\biggl\{-(i\hbar)^{-2}[\Delta,
\Psi]\biggr\}}S\exp{\biggl\{(i\hbar)^{-2}[\Delta,
\Psi]\biggr\}}.
\e{508}
This we view as  the natural
automorphism of the master equation \r{404}.
These transformations leave the
$\phi^a=\eta^a=0$ component of $S$ invariant
in a trivial manner since
$\left.[\Delta, \Psi(\phi, \eta)]\right|_{\phi=\eta=0}=0$.
The infinitesimal
invariance transformations and their properties
which follow from \r{508} are
\be
&&\del S=(i\hbar)^{-2}[S, [\Delta, \Psi]]=
(i\hbar)^{-2}\biggl( (S, \Psi)_{\Delta}-\half[\Delta, [\Psi,
S]]\biggr),\nn\\
&&\del_{21} S\equiv(\del_2\del_1-
\del_1\del_2)S=(i\hbar)^{-2}[S, [\Delta,
\Psi_{21}]]=(i\hbar)^{-2}\biggl(
(S, \Psi_{21})_{\Delta}-\half[\Delta, [\Psi_{21},
S]]\biggr),\nn\\
&&\Psi_{21}=(i\hbar)^{-2}(\Psi_2, \Psi_1)_{\Delta}.
\e{509}

Since $S=G$ is a trivial solution of the
master equation \r{404}, we find from
\r{508} the following expression
for the master charge $S$ in \r{401}
\be
&&S=\exp{\biggl\{-(i\hbar)^{-2}[\Delta,
\Psi]\biggr\}}G\exp{\biggl\{(i\hbar)^{-2}[\Delta,
\Psi]\biggr\}},
\e{510}
where $\Psi$ is the odd operator in \r{506}
which has the form \r{c10} in appendix B.
This is
 the general solution of the quantum master
equation \r{404} within the class
connected by the transformations \r{508}.
The transformations \r{508} act
transitivily on \r{510}.

\setcounter{equation}{0}
\section{Extended Lie equations}
The invariance transformations of the quantum
master equation \r{404} which follow
from the transformation formulas \r{501}
together with \r{505} suggest that we could
define extended group elements by (cf.\r{307})
\be
&&U(\phi, \eta)\equiv \exp{\{{i\over\hbar}
F(\phi,\eta)\}},\quad [\Delta, F]=0,\quad
[\tilde{G}, F]=0,
\e{601}
where $\tilde{G}$ is the extended ghost charge in \r{414}.
 $F(\phi,\eta)$ is
obviously given by \r{506}. Interestingly enough
$U(\phi)=\left.U(\phi,
\eta)\right|_{\eta=0}$ is equal to the general
solution of \r{304} and \r{311} which
provides for another argument in favor of the
quantum master equation. Now if we
also require $U(\phi, \eta)$ to be a unit
transformation at $\phi^a=\eta^a=0$ exactly
as in
\r{307} then the general solution is
\be
&&U(\phi, \eta)\equiv \exp{\biggl\{-(i\hbar)^{-2}[\Delta,
\Psi(\phi, \eta)]\biggr\}},
\e{602}
which at $\eta^a=0$ is identical to \r{308} if we make the
identification $\Psi(\left.\phi,
\eta)\right|_{\eta=0}=\rho(\phi)$. By means of \r{602}
we may then define transformed
states and operators by
\be
&&\vb\tilde{A}(\phi, \eta)|=\vb{A}|U^{-1}(\phi, \eta), \quad
\tilde{A}(\phi,
\eta)=U(\phi,
\eta){A}U^{-1}(\phi,
\eta),
\e{603}
which at $\eta^a=0$ are the group
transformed states and operators satisfying the Lie
equations \r{301}-\r{302} where
the operator connections
\r{304} have the form \r{311}.

Also \r{603}  are group transformed states and operators but
under the extended group elements \r{602} which are parametrized by the
supersymmetric pairs of parameters $\phi^a$ and $\eta^a$.
The states and operators in \r{603} satisfy the
extended Lie equations
\be
&&\vb
\tilde{A}(\phi,\eta)|\stackrel{\lea}{\tilde{D}}_a\equiv\vb
\tilde{A}(\phi,\eta)|\left(\stackrel{\lea}{\d_a}-(i\hbar)^{-1}
\tilde{Y}_a(\phi,\eta)\right)=0,\nn\\
&&\tilde{A}(\phi,\eta)\stackrel{\lea}{\tilde{\nabla}}_a\equiv
\tilde{A}(\phi,\eta)\stackrel{\lea}{\d_a}
-(i\hbar)^{-1}[\tilde{A}(\phi,\eta),
\tilde{Y}_a(\phi,\eta)]=0,
\e{607}
where
\be
&&\tilde{Y}_a(\phi,\eta)=i\hbar U(\phi, \eta)\left(U^{-1}(\phi,
\eta)\stackrel{\lea}{\d_a}\right).
\e{608}
The expression \r{602} for $U(\phi, \eta)$ implies then that
\be
&&\tilde{Y}_a(\phi,\eta)=(i\hbar)^{-1}
[\Delta, \tilde{\Om}_a(\phi,\eta)],
\e{609}
where
\be
&&\tilde{\Om}_a(\phi,\eta)=\nn\\&&=\int_0^1 d\al
\exp{\{-(i\hbar)^{-2}\al[\Delta,
\Psi(\phi,\eta)]\}}\left(\Psi(\phi,\eta)
\stackrel{\lea}{\d_a}\right)
\exp{\{(i\hbar)^{-2}\al[\Delta,
\Psi(\phi,\eta)]\}}.\nn\\
\e{610}

The general  expression \r{510} for
the master charge $S$ implies in particular
that the master charge $S$ itself is a
group transformed ghost charge under the
extended group element \r{602} which
means that it satisfies the Lie equation
\be
&&S(\phi,\eta)\stackrel{\lea}{\tilde{\nabla}}_a\equiv
S(\phi,\eta)\stackrel{\lea}{\d_a}-(i\hbar)^{-1}[S(\phi,\eta),
\tilde{Y}_a(\phi,\eta)]=0.
\e{611}
This equation together with the representation
 \r{609} of $\tilde{Y}_a(\phi,\eta)$
may be used to resolve $\tilde{\Om}_a(\phi,\eta)$
 in terms of $S$. We find from
\r{611} to the lowest order in $\eta^a$
\be
&&\tilde{\Om}_a(\phi,\eta)={\Om}_a(\phi)+
\half\eta^b\left(\Om_{ba}(-1)^{\ve_a}+(i\hbar)^{-1}[\Om_b,
\Om_a]\right)+\ldots
\e{612}
However, this solution is not unique. The arbitrariness is
\be
&&\tilde{\Om}_a\;\ra\;\tilde{\Om}_a+
(i\hbar)^{-1}[\Delta, \tilde{K}_a],
\e{6131}
which corresponds to  the arbitrariness
in \r{314} and \r{319}. (The $\eta^a\ra0$
limit yields \r{314}.) The relation between
the extended group transformation
and the nonextended one may also be expressed
in terms of the master charge $S$. We
have
\be
&&|\tilde{A}(\phi,\eta)\hb=
V(\phi,\eta)|{A}(\phi)\hb, \quad
\tilde{A}(\phi,\eta)=V(\phi,\eta){A}
(\phi)V^{-1}(\phi,\eta),
\e{614}
where
\be
 &&V(\phi,\eta)=U(\phi,\eta)U^{-1}(\phi).
\e{615}
Since
\be
&&i\hbar\al{d\over
d\al}U(\phi,\al\eta)=i\hbar\eta^a
{\d\over\d\eta^a}U(\phi,\al\eta)=[G,
U(\phi,\eta)]=(S(\phi,\al\eta)-G)
U(\phi,\al\eta),\nn\\
\e{616}
we get
\be
&&V(\phi,\eta)=T_{\al}\exp{\left\{(i\hbar)^{-1}\int_0^1
d\al{1\over\al}(S(\phi,\al\eta)-G)
\right\}}=1+(i\hbar)^{-1}\eta^a\Om_a+\nn\\&&+
(i\hbar)^{-1}{1\over4}\eta^a\eta^b\left(\Om_{ba}(-1)^{\ve_a}+
(i\hbar)^{-1}(\Om_{b}\Om_{a}+\Om_{a}
\Om_{b}(-1)^{(\ve_a+1)(\ve_b+1)})\right)+\ldots
\e{617}

\setcounter{equation}{0}
\section{Some further properties of
the extended states and operators}
Here we give some additional
properties and interpretations of the results in
section 6. First we notice that if the original states and
operators are eigenstates and
eigenoperators to the ghost charge $G$ then the
transformed states and operators in \r{603}
are eigenstates and eigenoperators to the
master charge $S$, \ie
\be
&&S|\tilde{A}\hb_g=i\hbar g|\tilde{A}\hb_g,
\quad [S, \tilde{A}_g]=i\hbar
g\tilde{A}_g.
\e{701}
Another
property which follows from the representation \r{603} is
\be
&&\Om|A\hb=0\quad\Rightarrow\quad\Delta|
\tilde{A}(\phi,\eta)\hb=0,\nn\\
&&[\Om, A]=0\quad\Rightarrow\quad[\Delta,
\tilde{A}(\phi,\eta)]=0,
\e{702}
\ie if the original states and operators are
invariant under the BRST charge $\Om$,
then the extended states and operators are
invariant under the extended BRST charge
$\Delta$.  $\Delta$ should of course be hermitian.

The extended BRST charge $\Delta$ may be
identified with the conventional BRST charge
if $\phi^a$ are identified with the
Lagrange multipliers and if $\xi_a$ and
$\eta^a$ are identified with antighosts and
their conjugate momenta. Although the
group element
$U(\phi,\eta)$ should be unitary, there
is an interesting interpretation of the
extended states
\be
&&|\tilde{A}\hb\equiv U(\phi,\eta)|A\hb,
\e{703}
when $U(\phi,\eta)$ is {\em hermitian}.
 If we choose $|A\hb$ to
be independent of the conjugate momenta
$\pet_a$ to the ghosts, \ie
\be
&&\ca^a|A\hb=0,
\e{704}
then $\Om|A\hb=0$ and $\Delta|A\hb=0$
which imply $\Delta|\tilde{A}\hb=0$. In fact,
the states in \r{703} are not only BRST
invariant under $\Delta$ but they are also
formal inner product states
\cite{BM}. A {\em hermitian} choice of $\Psi$
of the form $\Psi\propto \pet_a\phi^a$
in the representation \r{602} of $U$ in
\r{703} leads to inner product states with
group geometric properties. In \cite{Simple}
such states were considered for Lie group
theories.
\\ \\ \\
{\bf Acknowledgments}

I.A.B. would like to thank Lars Brink for
his very warm hospitality at the
Institute of Theoretical Physics, Chalmers
and G\"oteborg University.   The work of
I.A.B. is  supported by INTAS grant 96-0308
 and by RFBR grants 99-01-00980,
99-02-17916. I.A.B. and R.M. are thankful
to the Royal Swedish Academy of Sciences
for financial support.\\ \\ \\
\def\theequation{\thesection.\arabic{equation}}
\setcounter{section}{1}
\renewcommand{\thesection}{\Alph{section}}
\setcounter{equation}{0}
\noindent
{\Large{\bf{Appendix A}}}\\ \\
{\bf Quantum antibrackets.}\\ \\
Quantum antibrackets are algebraic tools
like commutators which should be useful in
any quantum theory in which there are
fundamental odd operators like in
BRST-quantization of gauge theories and
supersymmetric theories. So far they have
only been applied to gauge theories. They
provide for an operator formulation of the
BV-quantization \cite{Quanti,Sp2QA,GenQuanti}.
Here  as in \cite{OG,GenQuanti} they
are applied to the BFV-BRST quantization.

The basic quantum antibracket is defined by \cite{Quanti}
\be
&&(f, g)_Q\equiv\half \left([f, [Q, g]]-
[g, [Q, f]](-1)^{(\ve_f+1)(\ve_g+1)}\right),
\e{a1}
where  $f$ and $g$ are any operators with
Grassmann parities $\ve(f)\equiv\ve_f$ and
$\ve(g)\equiv\ve_g$ respectively. $Q$ is
an odd operator, $\ve(Q)=1$.  The commutators
on the right-hand side is the graded commutator \r{41}.
 The quantum antibracket \r{a1} satisfies the
properties:\\
\\ 1) Grassmann parity
\be
&&\ve((f, g)_Q)=\ve_f+\ve_g+1.
\e{a3}
2) Symmetry
\be
&&(f, g)_Q=-(g, f)_Q(-1)^{(\ve_f+1)(\ve_g+1)}.
\e{a4}
3) Linearity
\be
&&(f+g, h)_Q=(f, h)_Q+(g, h)_Q, \quad (\mbox{for}\; \ve_f=\ve_g).
\e{a5}
4) If one entry is an odd/even parameter $\la$ we have
\be
&&(f, \la)_Q=0\quad {\rm for\ any\ operator}\;f.
\e{a6}
5) the generalized Jacobi identities
(This general form was given in \cite{GenQuanti})
\be
&&(f,(g, h)_Q)_Q(-1)^{(\ve_f+1)(\ve_h+1)}+
cycle(f,g,h)=\nn\\&&={1\over
6}(-1)^{\ve_f+\ve_g+\ve_h}\left\{\left(
[f, [g, [h, Q^2]]]+\half[[f, [g,
[h, Q]]], Q]\right)(-1)^{\ve_f\ve_h}+\right.\nn\\
&&+\left.\left([f, [h,
[g, Q^2]]]+\half[[f, [h,
[g, Q]]],
Q]\right)(-1)^{\ve_h(\ve_f+\ve_g)}\right\}+cycle(f,g,h),
\e{a7}
6) and  the generalized Leibniz rule
\be
&&(fg, h)_Q-f(g, h)_Q-(f, h)_Qg(-1)^{\ve_g(\ve_h+1)}=\nn\\
&&=\half\left([f, h][g,
Q](-1)^{\ve_h(\ve_g+1)}+[f,Q][g,h](-1)^{\ve_g}\right).
\e{a8}
The properties 1)-4) agree exactly with the
corresponding properties of the classical
antibracket $(f,g)$ for functions $f$ and $g$.
However, the classical antibracket
satisfies in addition 6) and 7) where the
right-hand side is zero.
The right-hand side of \r{a8} is zero if we
confine ourselves to a maximal set of
commuting operators. They may \eg be
functions of a basic set of commuting coordinate
operators. The right-hand side of \r{a7}
is then zero if $Q$ and $Q^2$ are at most
quadratic in the conjugate momenta to these
coordinate operators. The classical
antibracket is then finally obtained
in the  coordinate representation.

The quantum antibracket \r{a1} is also
useful for arbitrary operators. However, in
this case we need further algebraic tools
represented by higher order quantum
antibrackets defined by
\cite{Quanti,GenQuanti}
\be
&&(f_{a_1},\ldots, f_{a_n})_Q\equiv -{1\over
n!}(-1)^{E_n}\sum_{\rm sym} [\cdots[[Q, f_{a_1}],
\cdots,f_{a_n}]\equiv\nn\\&&\equiv
-{1\over n!}[\cdots[[Q,
f_{b_1}],\cdots,f_{b_n}]\la^{b_n}\cdots\la^{b_1}
\stackrel{\lea}{\d}_{a_1}\stackrel{\lea}{\d}_{a_2}\cdots
\stackrel{\lea}{\d}_{a_n}(-1)^{E_n},\nn\\&&
\quad E_n\equiv
\sum_{k=0}^{\left[{n-1\over 2}\right]}\ve_{a_{\rm 2k+1}}.
\e{a9}
They satisfy the properties
\be
&&(\ldots, f_a, f_b, \ldots)_{Q}=
-(-1)^{(\ve_a+1)(\ve_b+1)}
(\ldots, f_b, f_a, \ldots)_{Q},\nn\\
&&\ve((f_{a_1},
f_{a_2},\ldots,f_{a_n})_{Q})=\ve_{a_1}+
\cdots+\ve_{a_n}+1.
\e{a10}
The higher order quantum antibrackets
may also be expressed recursively in terms of
the next lower ones. We have \cite{GenQuanti}
\be
&&(f_{a_1},\ldots, f_{a_n})_Q={1\over n}
\sum_{k=1}^n [(f_{a_1},\ldots,f_{a_{k-1}},
f_{a_{k+1}}, \ldots, f_{a_{n}})_Q, f_{a_k}](-1)^{B_{k,n}}, \nn\\
&&B_{k,n}\equiv\ve_{a_k}(\ve_{a_{k+1}}+\ldots
+\ve_{a_n})+\sum_{s=2[{k\over2}]+1}^n\ve_{a_s}.
\e{a11}
 To the lowest
order we have explicitly
\be
 &&(f_a, f_b, f_c)_Q={1\over
3}(-1)^{(\ve_a+1)(\ve_c+1)}\left([(f_a, f_b)_Q,
f_c](-1)^{\ve_c+(\ve_a+1)(\ve_c+1)}+cycle(a,b,c)\right).\nn\\
\e{a12}
The higher  quantum antibrackets  satisfy
the following generalized Jacobi identities
\cite{GenQuanti}
\be
&&\sum_{k=1}^n(f_{a_k}, (f_{a_1},\ldots,
f_{a_{k-1}}, f_{a_{k+1}}, \ldots,
f_{a_n})_Q)_Q(-1)^{D_{k,n}}=-{n-2\over2}[Q, (f_{a_1},\ldots,
f_{a_n})_Q]-\nn\\&&-R_n+(-1)^{E_n}{1\over n!}\sum_{\rm
sym}[\cdots[[Q^2,f_{a_1}], f_{a_2}],\cdots, f_{a_n}]=0,
\e{a13}
where
\be
&&R_n\equiv
\half\sum_{k=2}^{n-2}\sum_{\rm sym}[(f_{a_1},
\ldots, f_{a_k})_Q, (f_{a_{k+1}}, \ldots,
f_{a_n})_Q](-1)^{F_{k,n}},\nn\\&&F_{k,n}
\equiv \sum_{r=1}^{(n,k)}\ve_{a_r},
\e{a14}
where $(n,k)\equiv n$ for $k$ odd, and $(n,k)
\equiv k$ for $k$ even. The symmetrized
sum is over all different orders
with additional sign factors
$(-)^{E_n+\tilde{E}_n+A_n}$ where
$\tilde{E}_n$ is $E_n$ for the new order and $A_n$
from the reordering of the monomial
$\la^{a_1}\cdots\la^{a_n}$. For $n=3$ \r{a13}
reduces to \r{a7}.\\ \\ \\
\setcounter{section}{2}
\setcounter{equation}{0}
\noindent
{\Large{\bf{Appendix B}}}\\ \\
{\bf Proof of \r{506}.}\\ \\
Following section 5 we let  $F(\phi,\eta)$ be an operator which neither
depends on
$\pi_a$ nor
$\xi_a$ and which is  given by a power expansion in $\phi^a$
and $\eta^a$.
 We want to solve the condition \r{505}, \ie
\be
&&[\Delta, F(\phi,\eta)]=0.
\e{c1}
To this purpose we introduce the operators
\be
&&\Lambda\equiv \eta^a\pi_a(-1)^{\ve_a}, \quad
\tilde{\Lambda}\equiv
\xi_a\phi^a(-1)^{\ve_a},
\e{c2}
with the properties
\be
&&\ve(\Lambda)=\ve(\tilde{\Lambda})=1,\quad
{\Lambda}^2=\tilde{\Lambda}^2=0.
\e{c3}
$\Lambda$ has total ghost charge plus one and
$\tilde{\Lambda}$ minus one.
We have then
\be
&&\Delta=\Om+\Lambda, \quad N\equiv(i\hbar)^{-1}[\Lambda,
\tilde{\Lambda}]=\pi_a\phi^a+\xi_a\eta^a,\quad[N, {\Lambda}]=
[N,\tilde{\Lambda}]=0,
\e{c4}
from which it follows that
\be
&&(i\hbar)^{-1}[N, F(\phi,\eta)]=-\biggl(\phi^a{\d\over\d\phi^a} +
\eta^a{\d\over\d\eta^a}\biggr) F(\phi,\eta).
\e{c5}
By commuting \r{c1} with $\tilde{\Lambda}$
and taking \r{c4}, \r{c5} into account we
get
\be
&&\biggl(\phi^a{\d\over\d\phi^a} +
\eta^a{\d\over\d\eta^a}\biggr) F(\phi,\eta)
=(i\hbar)^{-2}[[F(\phi,\eta),
\tilde{\Lambda}], \Delta].
\e{c6}
Next, let us make the following canonical transformation in \r{c6}:
\be
&&\phi^a\;\ra\;\al\phi^a,\quad \pi_a\;\ra\;\al^{-1}\pi_a,\nn\\
&&\eta^a\;\ra\;\al\eta^a,\quad \xi_a\;\ra\;\al^{-1}\xi_a,
\e{c7}
where $\al$ is a parameter. Then we get
\be
&&\al{d\over d\al}F(\al\phi,\al\eta)=
(i\hbar)^{-2}[[F(\al\phi,\al\eta),
\tilde{\Lambda}], \Delta].
\e{c8}
Integration  over $\al$ yields then
\be
&&F(\phi,\eta)=F(0,0)+(i\hbar)^{-2}
[\int_0^1{d\al\over\al}[F(\al\phi,\al\eta),
\tilde{\Lambda}], \Delta]\equiv  F(0,0)+
(i\hbar)^{-1}[\Delta, \Psi],\nn\\
&&\left.[\Delta, \Psi]\right|_{\phi=\eta=0}=0,
\e{c9}
which is the assertion in \r{506}. Notice that
 $\Psi$ has the explicit form
\be
&&\Psi(\phi,\eta)=(i\hbar)^{-1}\int_0^1
{d\al\over\al}[F(\al\phi,\al\eta),
\tilde{\Lambda}] + (i\hbar)^{-1}[\Delta, K(\phi,\eta)],
\e{c10}
where $K(\phi,\eta)$ is an arbitrary
operator with total ghost number minus two
and which neither
depends on
$\pi_a$ nor
$\xi_a$. Notice that
\be
&&\left.\Psi(\phi,\eta)\right|_{\eta=0}=\int_0^1 d\al
F_a(\al\phi)(-1)^{\ve_a}+(i\hbar)^{-1}[\Om, K(\phi,0)],
\e{c11} where
$F_a(\phi)=\left.\biggl(F(\phi,\eta){
\stackrel{\lea}{\d}\over\d\eta^a}\biggr)\right|_{\eta=0}$.
\r{c11} agrees exactly with \r{3142} with the identifications
\be
&&\rho(\phi)=\left.\Psi(\phi,\eta)\right|_{\eta=0},\quad
G_a(\phi)=F_a(\phi)(-1)^{\ve_a},\quad \kappa(\phi)=K(\phi,0).
\e{c12}
The integrability conditions \r{3143}
of $G_a(\phi)$ are encoded in \r{c9}.
\newpage
\setcounter{section}{3}
\setcounter{equation}{0}
\noindent
{\Large{\bf{Appendix C}}}\\ \\
{\bf A generalization.}\\ \\
Consider the Lie equations \r{301}-\r{302}.
If we require the connection operator
$Y_a$ to only satisfy $[\Om, Y_a]=0$, then we have
\be
&&Y_a=Y^{(S)}_a+(i\hbar)^{-1}[\Om, \Om_a],
\e{b1}
where $Y^{(S)}_a$ is the BRST singlet part of $Y_a$.
The integrability conditions \r{303} imply then
\be
&&Y^{(S)}_a\stackrel{\lea}{\d_b}-Y^{(S)}_b
\stackrel{\lea}{\d_a}(-1)^{\ve_a\ve_b}-
(i\hbar)^{-1}[Y^{(S)}_a,
Y^{(S)}_b]+ \nn\\&&+(i\hbar)^{-1}[\Om,
\left(\Om_a\stackrel{\lea}{\d_b}-
(i\hbar)^{-1}[\Om_a,Y^{(S)}_b]\right)
-\nn\\&&-\left(\Om_b\stackrel{\lea}{\d_a}(-1)^{\ve_a\ve_b}-
(i\hbar)^{-1}[\Om_b,Y^{(S)}_a]\right)(-1)^{\ve_a\ve_b}-
(i\hbar)^{-2}(\Om_a,
\Om_b)_{\Om}]=0.
\e{b2}
These conditions are natural to split as follows
\be
&&Y^{(S)}_a\stackrel{\lea}{\d_b}-Y^{(S)}_b
\stackrel{\lea}{\d_a}(-1)^{\ve_a\ve_b}-
(i\hbar)^{-1}[Y^{(S)}_a,
Y^{(S)}_b]=0,\nn\\
&&\Om_a\stackrel{\lea}{\nabla^{(S)}_b}
-\Om_b\stackrel{\lea}{\nabla^{(S)}_a}(-1)^{\ve_a\ve_b}-
(i\hbar)^{-2}(\Om_a,
\Om_b)_{\Om}=-\half(i\hbar)^{-1}[\Om_{ab}, \Om],
\e{b3}
where
\be
&&\stackrel{\lea}{\nabla^{(S)}_a}\equiv
\stackrel{\lea}{\d_a}-(i\hbar)^{-1}[\;\cdot\;,Y^{(S)}_a],\quad
[\stackrel{\lea}{\nabla^{(S)}_a}, \stackrel{\lea}{\nabla^{(S)}_b}]=0.
\e{b4}
Since the Lie equations \r{301}-\r{302} may  be written as
\be
&&\vb
A(\phi)|\left(\stackrel{\lea}{\nabla^{(S)}_a}-
(i\hbar)^{-1} Y_a(\phi)\right)=0, \quad
A(\phi)\stackrel{\lea}{\nabla^{(S)}_a}-
(i\hbar)^{-1}[A(\phi), Y_a(\phi)]=0,
\e{b5}
where $Y_a(\phi)\equiv(i\hbar)^{-1}[\Om, \Om_a]$,
it follows that we still have the
quantum master equation \r{404} but here with
the $\Delta$-operator defined by
\be
&&\Delta\equiv\Om+\eta^a(\pi_a-Y^{(S)}_a)(-1)^{\ve_a}.
\e{b6}
In this case we find the general group element
$\tilde{U}(\phi,\eta)=U^{(S)}(\phi,\eta)U(\phi,\eta)$.
 $Y^{(S)}_a$ is expressed in
terms of $U^{(S)}(\phi,\eta)$ and $U(\phi,\eta)$
is given by \r{603} in section 6.
Thus, the singlet part of $Y_a$ may be transformed away.

\end{document}